\documentclass[]{aa}

\newcommand{\avg}[1]{\ensuremath{\langle #1 \rangle}}

\begin{document}
\title{3D Radiative MHD Simulations of Starspots II: Large-scale Structure}
\author{Tanayveer Singh Bhatia\inst{1}\fnmsep\thanks{Corresponding author: bhatia@mps.mpg.de}
        \and Mayukh Panja\inst{1}
        \and Robert H. Cameron\inst{1}
        \and Sami K. Solanki\inst{1}
        }
\institute{Max Planck Institute for Solar System Research, Justus-von-Liebig Weg 3, D-37077, Göttingen, Germany\\
           \email{bhatia@mps.mpg.de}}
\date{Received XXXX}

\abstract
{We compute realistic 3D radiative MHD near-surface models of starspots with substantial penumbrae on cool main-sequence stars using the MURaM simulation code. This work is an improvement on the the previous starspot models in a slab geometry. The umbra, penumbra and the quiet star for all starspots are distinct, not only in intensity and temperature, but also in thermodynamic and velocity structure. These models represent a significant step towards modelling contribution of starspots to stellar lightcurves.}

\maketitle
\begin{nolinenumbers}
\section{Introduction} \label{sec:intro}

Sunspots are the most prominent visible aspect of the Sun's 11-year magnetic cycle, with a vast amount of literature dedicated to their study \citep{solanki2003,scharmer2009,spotmagrev2011}. On the other hand, their stellar counterparts, starspots, are comparatively less well studied \citep{berdyugina2005,strassmeier2009}. On the sun, spots are quite dynamic, inhomogeneous and display considerable fine structure, with a dark core with suppressed convection called the umbra and a comparatively brighter filamentary outer component called the penumbra. The umbra harbours small-scale brightenings called umbral dots \citep{solanki2003} which correspond to convective flows. The penumbra is composed of alternating bright and dark components  \citep{solanki2003} and exhibits a radially\footnote{Unless otherwise stated, "radial" in the context of simulation results as well as sunspot observations refers to a polar coordinate system in the plane of the surface, with the spot center as the origin} outward flow called the Evershed flow \citep{evershed1909}. Spots appear geometrically depressed towards their center, an effect termed Wilson depression (WD) \citep{wilson1774}.

There are a number of indications that basically all stars with an external convection zone exhibit some form of magnetism \citep{baliunas1985,baliunas1995}. Many of these stars also appear to harbour starspots \citep{strassmeier2009}. The fact that we can observe most stars only as a point source makes it understandably difficult to study the detailed properties of starspots. Nevertheless, starspots affect photometric variability as well as measurements of radial velocities (RV) of stars. The radial velocity technique is used to detect and study properties of exoplanets \citep{rvstate2016,rvrev2023}. In addition, spots contribute to the transmission spectrum of exoplanets (spectrum of stellar light modified by passing through the exoplanetary atmosphere) and complicate their detection and characterization \citep{transspec2023}.

To understand the impact of starspots on stellar spectra, the spots are usually modelled as a 1D radiative equilibrium model atmosphere of some specific effective temperature \citep{soap2012,soap2014}. If starspots are anything like sunspots, that assumption is extremely simplified and introduces systematic errors in inferring details about exoplanets from stellar spectra. In fact, stellar variability is currently the single largest barrier to achieving greater precision in RV observations \citep{eprv2021}.

Simulations of sunspots reproduce the essential characteristics of actual sunspots, including penumbra and Evershed flow \citep{spotsimrev2011} (but see, e.g. \citet{jurcak2020}). These simulations indicate that penumbral filaments are essentially the result of convection in a highly inclined magnetic field geometry \citep{rempel2012}. In this paper, we extend such sunspot simulations to stars of other spectral types. Specifically, we present first simulations of starspots on a K2V and an M0V star, along with a reference G2V spot, in circular geometry. This paper is an extension to the slab geometry starspots modelled in \citet{paper1spots}.

The paper is organized as follows. In Section \ref{sec:methods} (as well as Appendix \ref{app:intensity}), we give an overview of the simulation setup and data analysis. In Section \ref{sec:results}, we show the average radial intensity, magnetic field and velocity, as well the the thermodynamic structure. In Section \ref{sec:discuss}, we discuss the physical implications of our simulations, and in Section \ref{sec:summary}, we summarize our work and present an outlook.

\section{Methods} \label{sec:methods}

The simulations presented here were computed using the MURaM radiative magnetohydrodynamic (MHD) code \citep{muram1,muram2}. For each starspot, the following procedure was carried out to set up the simulations: Hydrodynamic boxes using a mean stratification from earlier simulations (in this case, from \citet{paper1ssd}) were set up with the dimensions mentioned in Table \ref{tab:setup} but with half the resolution. The vertical extent of the boxes are such that there are roughly an equal number of pressure scale heights below the surface ($\ln(p/p_0) \sim 8$). The number of pressure scale heights is calculated for the pressure stratification in the quiet star region (with the quiet star region chosen as defined in Appendix \ref{app:intensity}). The horizontal extent for the G2V star was fixed to 48 Mm in both directions, and the rest scaled with the number of pressure scale heights to have roughly similar resolution for granules in each box. At this stage, the opacities used for the radiative transfer were gray (that is, no wavelength dependence). The FreeEOS equation of state was used for all runs \citep{freeeos}.

To initialize, the simulation boxes were seeded with a random very low-magnitude velocity field, which naturally developed into turbulent convection over time. Once the convection reached a dynamic equilibrium, a flux tube with 3 kG photospheric field and 30 kG field at the bottom boundary, with the photospheric spot radius $\sim 0.3 L_X$, where $L_X$ is the horizontal extent of the box in the $x$ direction (with $x,y$ as the horizontal directions and $z$ as the vertical direction). We followed the same procedure as in \citet{paper1spots} for setting up the spots. Briefly, the vertical field is set to be uniform in the horizontal plane and decaying exponentially in the vertical direction with a scale height determined from field strength specified at the surface and the bottom boundary. The horizontal field is radially symmetric and is such that $\nabla \cdot B = 0$. The radial extent of the spot is set to conserve the flux at the surface. In addition, the edges of the flux tube were enhanced with an additional ring of flux with 2 kG photospheric field strength and 10 kG field strength at the bottom boundary. This was done to promote the formation of a significant penumbra \citep{paper2spots}. The thickness of this ring at the surface was set to about half the diameter of the main flux tube at the surface. Lastly, to promote the formation of a stable penumbra, the horizontal magnetic field at the top boundary was enhanced (compared to a potential field extrapolation) using the approach described in \citet{rempel2012}. The corresponding value of $\alpha$ in our setup was 1.5.

The simulations were run for a few hours, till the initial transient vanished and the spot stabilized. Thereafter, we switched from gray opacities to 4-bin opacities based on the opacity distribution function (ODF) approach \cite{kurucz1993}. An ODF approach is routinely used in 3D stellar simulations as a compromise between numerical accuracy and computational cost \citep{stagger1,beeck1,witzke2024} The bin were split in opacity at $\log_{10}\tau_{500} = -0.5,-1.5,-3$. The initial reference model was chosen to be the horizontally averaged respective starspot with gray opacities. The simulations were run for another half an hour or so of stellar time, after which the opacity binning procedure was iterated with the new atmosphere. This procedure was repeated twice to ensure consistency. Lastly, the resolutions were doubled to the ones listed in Table \ref{tab:setup}.

The analysis presented in the subsequent sections is over one hour of stellar time for all simulations. The umbral and penumbral boundaries are determined semi-automatically based on the intensity histograms and visual appearance (see \ref{app:intensity} for details). The azimuthal averages are done by constructing pixel-wide ring-shaped masks, centered at the spot center, with increasing radius and taking the horizontal average in each of these masks. For averages of $\tau=1$ surfaces, this results in line plots. For averages of the simulation cubes, this results in a 2D plot. The former are averaged over 10 snapshots each and the latter are averaged over 6 snapshots each, with approximately constant cadence. All horizontal averages (in $x$ and $y$ direction) for a quantity $q$ are denoted by angular brackets $\avg{q}$, unless otherwise stated.

\section{Results} \label{sec:results} 

\begin{table}[h!]
\caption{\label{tab:setup}Simulation parameters for stellar spots}
\centering
\resizebox{\columnwidth}{!}{%
\setlength{\tabcolsep}{2pt} 
\begin{tabular}{l|cccccc}
\hline\hline
Star & $L_X,L_Z$ & $dx,dz$ & $g_{\rm surf}$ & $T_{\rm qs}$ & $T_{\rm p}/T_{\rm qs}$ & $T_{\rm u}/T_{\rm qs}$ \\
 &  (Mm) &  (km) &  (cm/s$^2$) &  (K) &  &  \\
\hline
G2V & 48, 4.50 & 46.9, 15.6 & $2.74 \times 10^4$ & 6092 $\pm$ 8 & 0.89 & 0.70 \\
K2V & 30, 2.88 & 29.3, 9.80 & $4.06 \times 10^4$ & 4856 $\pm$ 4 & 0.93 & 0.83 \\
M0V & 12, 1.07 & 11.72, 3.92 & $6.70 \times 10^4$ & 3858 $\pm$ 1 & 0.97 & 0.89 \\
\hline
\end{tabular}%
}
\tablefoot{\small The boxes have the same extent and resolution in the $X$ and $Y$ horizontal directions. The temperatures listed are the average disc-center effective temperature calculated from the bolometric intensity (see Appendix \ref{app:intensity} for further details). $L_X,L_Z$ refer to the horizontal ($L_X=L_Y$) and vertical extent of the simulation boxes, $dx,dz$ refer to the horizontal ($dx=dy$) and vertical resolution, $g_{\rm surf}$ refers to the constant surface gravity, and the subscripts $qs,u,p$ refer to the quiet-star, umbral, and penumbral temperatures respectively, with umbra, penumbral, and quiet star region as described in Appendix \ref{app:intensity}.}
\end{table}

\subsection{Surface structure}

\begin{figure*}[h!]
    \centering
    \includegraphics[width=17cm]{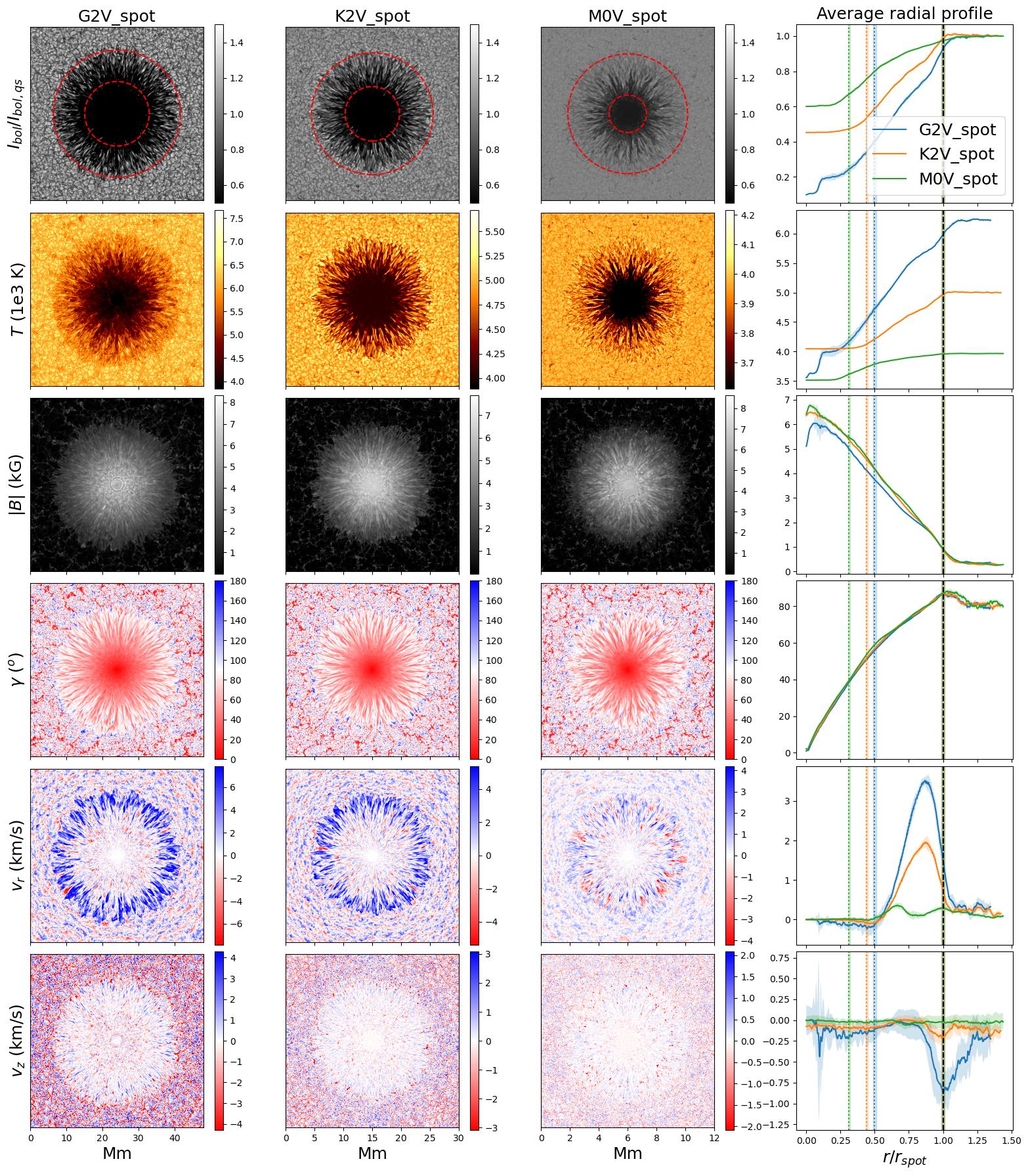}
    \caption{Snapshots of surface ($\tau=1$) structure of bolometric intensity (\textit{first row}), temperature (\textit{second row}), magnetic field magnitude (\textit{third row}), field inclination (\textit{fourth row}), radial velocity (\textit{fifth row}) and vertical velocity (\textit{sixth row}) for the G2V (\textit{first column}), K2V (\textit{second column}) and M0V (\textit{third column}) starspot. The rightmost (\textit{fourth}) column shows the azimuthal average of the corresponding quantities in each row for all three stars, plotted against radius normalized by average spot radius. The dashed lines show the average extent of the penumbra and dotted lines show the average extent of the umbra. These extents are shown in the first row as red circles. The error bars show 1$\sigma$ standard deviation averaged over time. For $v_z$, positive values correspond to upwards/outwards flow.}
    \label{fig:surface}
\end{figure*}

All spot simulations showcase spot structure that appear qualitatively similar to sunspots, with a dark umbra and a comparatively lighter penumbra composed of penumbral filaments, inclined fields and an Evershed flow.

In Figure \ref{fig:surface}, we plot the emergent bolometric intensity normalized by the horizontally averaged value ($I_{\rm bol}/\avg{I_{\rm bol}}$, where $\avg{I_{\rm bol}}$ is averaged over the quiet star region as defined in Appendix \ref{app:intensity}), as well as the temperature ($T$), magnetic field strength ($B$), magnetic field inclination ($\gamma = {\rm tan}^{-1}(B_h/B_z)$), radial velocity ($v_r$) and vertical velocity ($v_z$) at the $\tau=1$ iso-surface \footnote{"iso-surface" here refers to the corrugated surface formed from the height of each vertical column in the simulation cube where $\tau$ approaches 1.}. In addition, in the rightmost column, we also plot the azimuthal and time-averages of all the aforementioned quantities against radius normalized by the average spot radius for each stellar type. We note that the umbral-penumbral boundary (see Appendix \ref{app:intensity} for details) is somewhat smaller for the M0V spot, compared to the G2V and the K2V spot.

In the intensity plot (first row), we see that the contrast between the umbral and quiet star intensity decreases with decreasing $T_{\rm eff}$. In addition, the penumbra in M0V spot looks qualitatively different from the G2V and the K2V penumbra. For example, in the M0V spot, the outer penumbral structure is almost the same intensity as the quiet star region, the only distinguishing feature being the elongated penumbral filaments, whereas for the other two cases, the full extent of the penumbra is visibly darker than the quiet star photosphere. We list the average umbral and penumbral contrasts in Table \ref{tab:setup} for reference. We see the same trend reflected in the temperature plots as well. These trends are well in line with slab geometry spots in \citet{paper1spots}, which implies the geometry of spots is not a crucial factor in determining spot properties. As was noted for the slab spots, this trend in contrast is due to a strong dependence of the largest source of opacity, $\rm H^{-}$ ions, on temperature in the range of temperatures considered here.

The magnetic field shows a remarkably consistent trend with spot radius between the stellar types, with the coolest star having somewhat stronger fields overall, but not by much (less than 1 kG). As explored in \citet{paper1spots}, the spot field strength is determined not just by the quiet star gas pressure, but also the WD. We will discuss this further in Section \ref{sec:discuss}. The field inclination also follows a remarkably similar trend, despite the umbral-penumbral boundary being at different locations for different stellar types.

The radial velocity plots show that all three cases develop an outward Evershed-like radial flow. Velocities are larger for hotter stars. All spots also show inflow channels that are sometimes observed in sunspot penumbrae \citep{sebas2021}. The inflow channels, or counter Evershed flows, seem to be more frequent for the coolest star, but the radial average still shows a net outward flow. The magnitude of Evershed flow is probably related to the strength of the Lorentz force term relative to the advective term in the MHD momentum equation. Relatively stronger fields dictate where flows go and vice versa. Lastly, maps of vertical velocity show a average downflow at the spot boundaries, the magnitude of which seems to decrease with $T_{\rm eff}$. These downflows are observed for sunspots as well, and are believed to be the returning component of the Evershed flow \citep{michiel2013}. We note that, even though the vertical velocity structure at the optical can seemingly imply non-conservation of mass, this is not actually the case. At any given geometrical height, mass flux is conserved. The optical surface, on the other hand, is geometrically deeper in the umbral and penumbral region, compared to the quiet star region, and any inference of mass conservation needs to be considered more carefully (see Sec. \ref{sec:res:conv}).

\subsection{Convective structure} \label{sec:res:conv}

\begin{figure*}[ht]
    \centering
    \includegraphics[width=17cm]{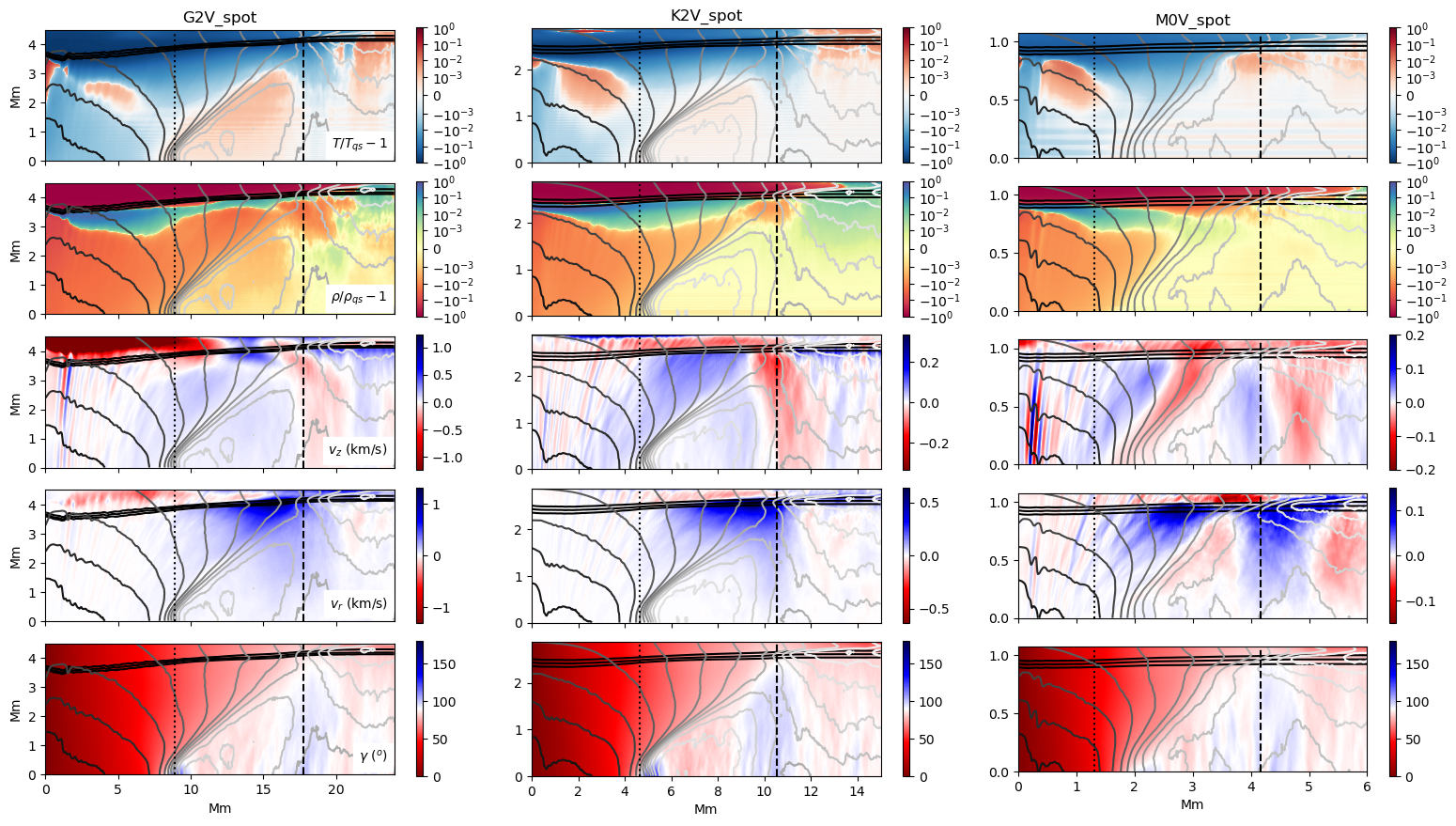}
    \caption{Time averages of azimuthally averaged relative change in temperature $T/T_{\rm qs} -1$ (\textit{first row}), relative change in density $\rho/\rho_{\rm qs} -1$ (\textit{second row}), vertical velocity (\textit{third row}), radial velocity (\textit{fourth row}), and magnetic field inclination (\textit{fifth row}) for the G2V (\textit{first column}, K2V (\textit{second column}) and M0V (\textit{third column}) starspot. The spot center corresponds to zero on the horizontal axis. The black contour lines near the top boundary represent the $\log_{10}\tau = -1,0,1$ levels, from top to bottom, respectively. The grey contour lines represent regions of similar magnetic field strength ranging from $\log_{10}(B/B_{\rm max} = $ -2 to 0 in 15 levels, where $B_{\rm max}=$ 14.88, 19.46 and 20.13 kG for the G2V, K2V and M0V starspot, respectively. The colormaps of the first two plots are symmetric log, to represent the large variation that can exist in temperature and density inside and outside of a starspot. To showcase the variation better, the vertical scale is expanded relative to the horizontal scale}
    \label{fig:convection}
\end{figure*}

We examine the azimuthally averaged thermodynamic and convective structure in Figure \ref{fig:convection}. The first two rows show the average temperature and density, relative to the quiet star (with quiet star region chosen as described in Appendix \ref{app:intensity}) temperature and density, respectively. These are, accordingly, plotted as $T/\avg{T_{\rm qs}}-1$ and $\rho/\avg{\rho_{\rm qs}}-1$, to show where there is an excess or deficit, relative to the quiet star structure. The umbral region right below the surface (at $\sim 3.5$ Mm for the G2V spot and in similar relative location for other spots) as well as near the bottom boundary show a relative decrease of temperature around a few percent (blue-ish region in the first row). The former also corresponds to a region of relatively higher density (green-ish region in second row). Just below this region is a region with somewhat higher temperature compared to the quiet star temperature (at $\sim 2.5$ Mm for the G2V spot and in a similar relative location, but stronger and over a wider region, for the other spots). We also see trends of a hot ring in the subsurface penumbral region starting somewhere in the middle of the box and extending all the way to the bottom (well visible for the G2V star, but much weaker for the K2V spot and basically not visible at all for the M0V spot).

In the velocity structure, we see a broad mean upflow (third row) and outflow (fourth row) surrounding the umbral trunk, and a downflow on the periphery of the penumbra. These trends qualitatively match those seen in sunspot simulations of \citet{rempel2015}. In addition, for the M0V star, between these two flows there is an additional downward (inward) and upward (outward) flow. The K2V spot also shows hints of these additional flows, especially near the bottom boundary, but these are all but invisible for the G2V spot. All these trends point to a ringed convective pattern around the spot region, with magnetic field inclination affecting the angle at which convective plumes rise and fall. We discuss this further in Sec. \ref{sec:discuss}.

Lastly, in the fifth row, we see that the azimuthally averaged field inclination for all spots points to a largely vertical field in the umbral trunk, that slowly gets more horizontal in the penumbra, even at depth. Right at the edge of the penumbra, there is region of field of opposite polarity extending all the way to the bottom. Based on the magnetic field strength contours, this possibly points to submergence of horizontal field at the edge of penumbra through regular convection.

\section{Discussion} \label{sec:discuss}

\begin{figure}[h]
    \centering
    \includegraphics[width=\linewidth]{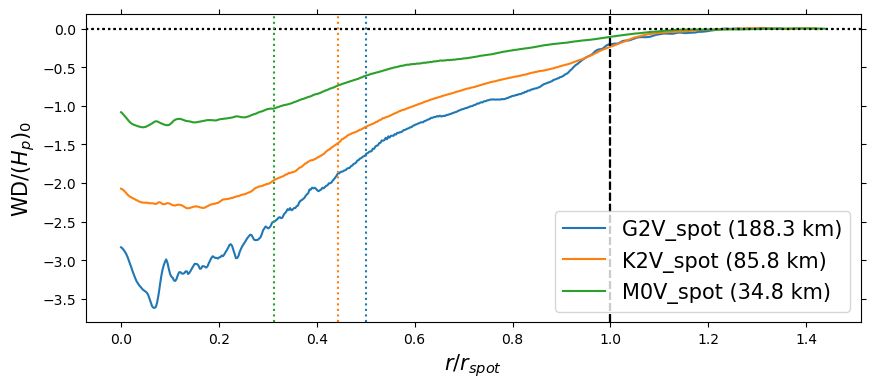}
    \caption{Wilson depression, normalized by the average quiet star pressure scale height $H_{p}$ at the $\tau=1$ level, plotted against normalized spot radius. The colors refer the same stellar types as in Figure \ref{fig:surface}.}
    \label{fig:wd}
\end{figure}

The round starspot simulations are consistent with the trends in \citep{paper1spots}, that is, cooler starspots have lower intensity and temperature contrast as well as smaller flow velocities. We see in Table \ref{tab:setup} that the brightness temperature contrast between the umbra and quiet star is higher (temperature ratio lower) than the penumbral contrast. This is to be expected and is consistent with observations where contrasts obtained using molecular spectra (sensitive to umbra) are higher than those obtained from modelling light curves \citep{oneal2004}.

The variation of field strength and field inclination with spot radius is quite similar for all cases. Naively, one would expect field strength to scale with average gas pressure, in which case the M2V star should have the strongest fields. But this dependence is complicated by the relative height at which the spot is observed, which is lower in the spot compared to the surroundings due to the WD. From the iso-tau contours in Figure \ref{fig:convection}, one sees a dip in the surface near where the spot is, before levelling off in the quiet star region. In Figure \ref{fig:wd}, we plot the average WD, normalized by the quiet star pressure scale height at the surface, as a function of normalized spot radius for all the stars. Here we see that the WD is larger for the hotter starspots relative to the pressure scale height, which implies that the field is observed deeper down in the atmosphere. This deeper field is in horizontal pressure balance with relatively higher pressure, and hence has a larger magnitude that expected from horizontal pressure balance against surface gas pressure.

\begin{figure}[ht]
    \centering
    \includegraphics[width=\linewidth]{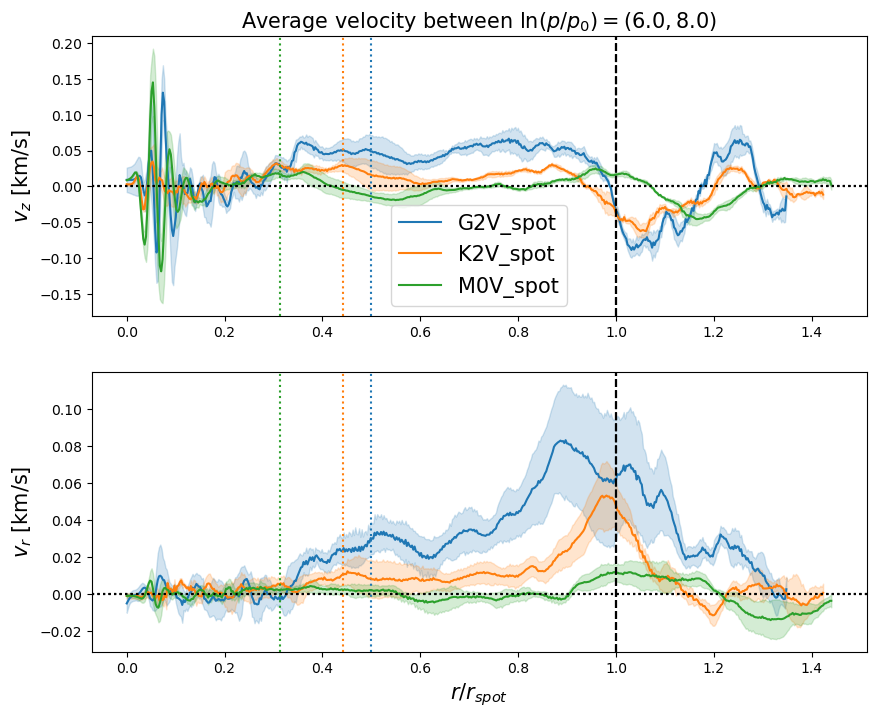}
    \caption{The vertical velocity $v_z$ (\textit{top}) and the radial velocity (\textit{bottom}) averaged between $\ln {p/p_0} = $ 6 and 8. The error bars indicate the 1$\sigma$ standard deviation in the averaging, from the data in Fig. \ref{fig:convection}. The horizontal axis and the dashed and dotted lines are the same as in the previous figures.}
    \label{fig:velocity}
\end{figure}

\begin{figure}[ht]
    \centering
    \includegraphics[width=\linewidth]{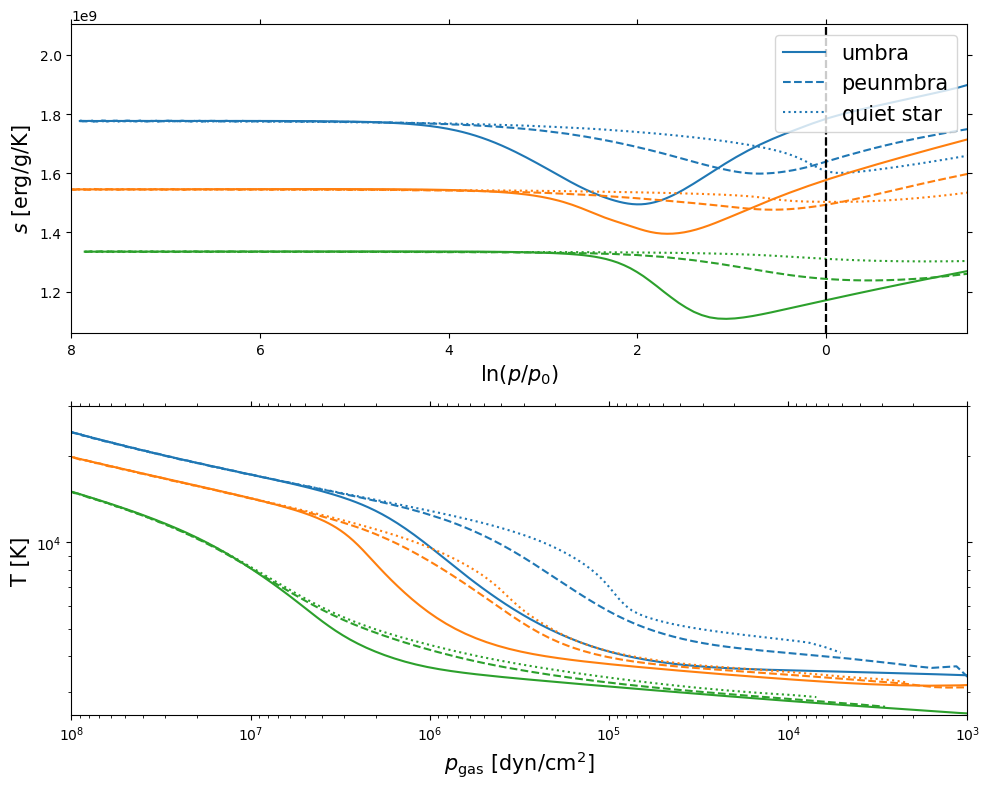}
    \caption{\textit{Top:} Average entropy in the umbral (\textit{solid}), penumbral (\textit{dashed}) and quiet star (\textit{dotted}) region plotted against pressure in units of the pressure scale height $\ln(p/p_0)$ calculated from the pressure stratification in the quiet star region. \textit{Bottom:} The temperature stratification for all the cases plotted against the gas pressure in logscale. The dashed vertical line represents the stellar surface. The bottom boundary is located towards the left side of the plot. The colors refer to the same stellar types as in Figure \ref{fig:surface}.}
    \label{fig:entropy_str}
\end{figure}

The major effect of the spot field locally is to block heat flux by reducing the efficiency of convection. This forces convection to have a more radial pattern. To this effect, we see an indication of a hot ring near the bottom boundary corresponding to the penumbral region for the G2V spot, but no such signature for the K2V and the M2V spot. Higher up, we do see a hint of an additional hot ring just outside the penumbra for the M2V spot, which would further support the convective structure being ringed. To confirm this, we plot $v_r$ and $v_z$ averaged near the bottom of the box in Fig. \ref{fig:velocity}. The flow structure clearly indicates a ringed convective structure, with the magnitude decreasing with $T_{\rm eff}$. In addition, in the intensity profile, we see no clear indication of a bright ring surrounding the spot in the bolometric intensity. These results are qualitatively consistent with previous simulations of sunspots \citep{rempel2011}. The thermodynamic stratification in the umbral trunk differs significantly from the corresponding quiet star stratification for all cases. This can be understood in terms of changes in the criteria for convection in the presence of strong magnetic fields. Instead of the usual Schwarzschild criterion for convection \citep{schwarzschild}, the Gough-Tayler criterion must be satisfied \citep{gt1966} in the presence of a strong  vertcal magnetic field. Briefly, in the presence of a vertical magnetic field, the requirement for convection becomes
\begin{equation} \label{eq:gt}
    \nabla - \nabla_{ad} - \frac{B_z^2}{B_z^2 + 4\pi \Gamma p} < 0
\end{equation}
Here, $\Gamma$ is the adiabatic index, $p$ is the gas pressure, $B_z$ is the vertical field, $\nabla = d\ln T/d\ln p$ is the temperature gradient, and the subscript $ad$ refers to the adiabatic temperature gradient $\nabla_{ad}=1-1/\Gamma$. When $B_z=0$, Eq.\ref{eq:gt} reduces to the Schwarzschild criterion. The applicability of this criterion was recently studied by \citet{schmassmann2021}, where they show that umbral, penumbral and quiet sun convection are distinct regimes of convection based on this criterion. Even though sunspots clearly do not have a simple uniform vertical field geometry, it is instructive to try and understand the effect of fields on convection in this frame of reference.

The umbra is largely free of convection, except for the presence of umbral dots, which are driven primarily by radiative cooling at the surface. Near the surface  where $B_z$ becomes comparable to $p_{\rm gas}$, the last term on the left hand side of Eq. \ref{eq:gt} becomes quite significant and the character of convection changes from regular somewhat field-free convection to strongly magnetized umbral dots-like convection in a more-or-less stable stratification. This is reflected in the entropy profile as well (Fig. \ref{fig:entropy_str}, top panel), where the gradient in the umbral entropy becomes positive below the surface (at $\ln(p/p_0) \sim 2-3$ for the G2V spot) signalling stable stratification. Since the interface between turbulent convection and stable stratification near the surface is marked by a steeper temperature gradient and a flatter density gradient due to a transition from convective to radiative energy transport (see, e.g., Figure 10 of \citet{beeck1}), the changes in umbral density and temperature profile can largely be understood as field-induced stable stratification near the surface leading to suppression of convection below. In general, the stratification for the umbra, penumbra and quiet-star region is significantly different near the surface (Fig. \ref{fig:entropy_str}, bottom panel). This indicates that stellar spots not only have an effect on the emergent intensity, but also affect structure quite significantly. Most commonly used strategies for modelling starspot signal in stellar spectra use simplified approaches that assume spots as a uniform circular feature with either some black-body profile \citep{spotrod,soap,pytranspot} or a 1D model atmosphere with a different effective temperature \citep{starsim}. The models presented here and the corresponding synthetic spectra (Smitha et. al., submitted ApJL) can be used for more realistic limb darkening laws and spot contrasts for modelling variations in stellar light curves.

\section{Summary and Outlook} \label{sec:summary}

The basic structure of a starspot is qualitatively similar across all the models considered here, namely, a dark umbra, surrounded by a relatively brighter filamentary penumbra harbouring a radially outward flow. However, the assumption of a homogeneous spot with a single effective temperature assumed in most stellar models incorporating spots for constructing light curves is oversimplified. The temperature and density structure cannot be simply modeled as a 1D radiative equilibrium stellar model. The large-scale flow structures have sufficient velocity to affect RV signals depending on where the spot is on the stellar disk. We investigate the effect of spots on synthetic spectra in a separate publication \citep{smitha_spot_spectra}.

We note that the presented models are somewhat exploratory in nature. Our setup is such that all models start with the same initial flux tube, scaled by pressure scale height. Since stellar radii don't scale as strongly with stellar type as pressure scale height, this in practice implies proportionally smaller spots on cooler stars. This is in contrast to observations, where starspots on cooler (and usually more active) stars are expected to cover a larger surface area. Even on the Sun, there are correlations between umbral temperature on size and field strength (see Section 5.7 of \citet{solanki2003}). With that being said, these models represent a step towards a more realistic accounting of effect of spots on stellar structure and observations. Prior tests in 2D geometry for a range of field strengths at the bottom boundary showed only a weak dependence of intensity and surface field strength on field at the bottom boundary (see Appendix of \citet{paper1spots}). Finally, we note that all realistic radiative MHD simulations of sun/starspots work in the monolithic tube paradigm. For future work, we plan to study the effect of spot size and geometry on various characteristics studied here and extend the grid of spot models to other spectral types.

\begin{acknowledgements}
The authors acknowledge useful comments from the anonymous referee that helped in improving this manuscript. TB acknowledges Damien Przybylski and Veronika Witzke for helping out with the equation of state. TB is also grateful for access to the supercomputer Cobra at Max Planck Computing and Data Facility (MPCDF), on which all the simulations were carried out. This project has received funding from the European Research Council (ERC) under the Advanced Grant scheme (grant agreement No. 101097844 - project WINSUN).
\end{acknowledgements}

\bibliography{bib}{}
\bibliographystyle{aa}

\begin{appendix}
\onecolumn
    \section{Intensity contours and averaging}\label{app:intensity}
    \begin{figure}[h]
    \centering
    \includegraphics[width=0.75\linewidth]{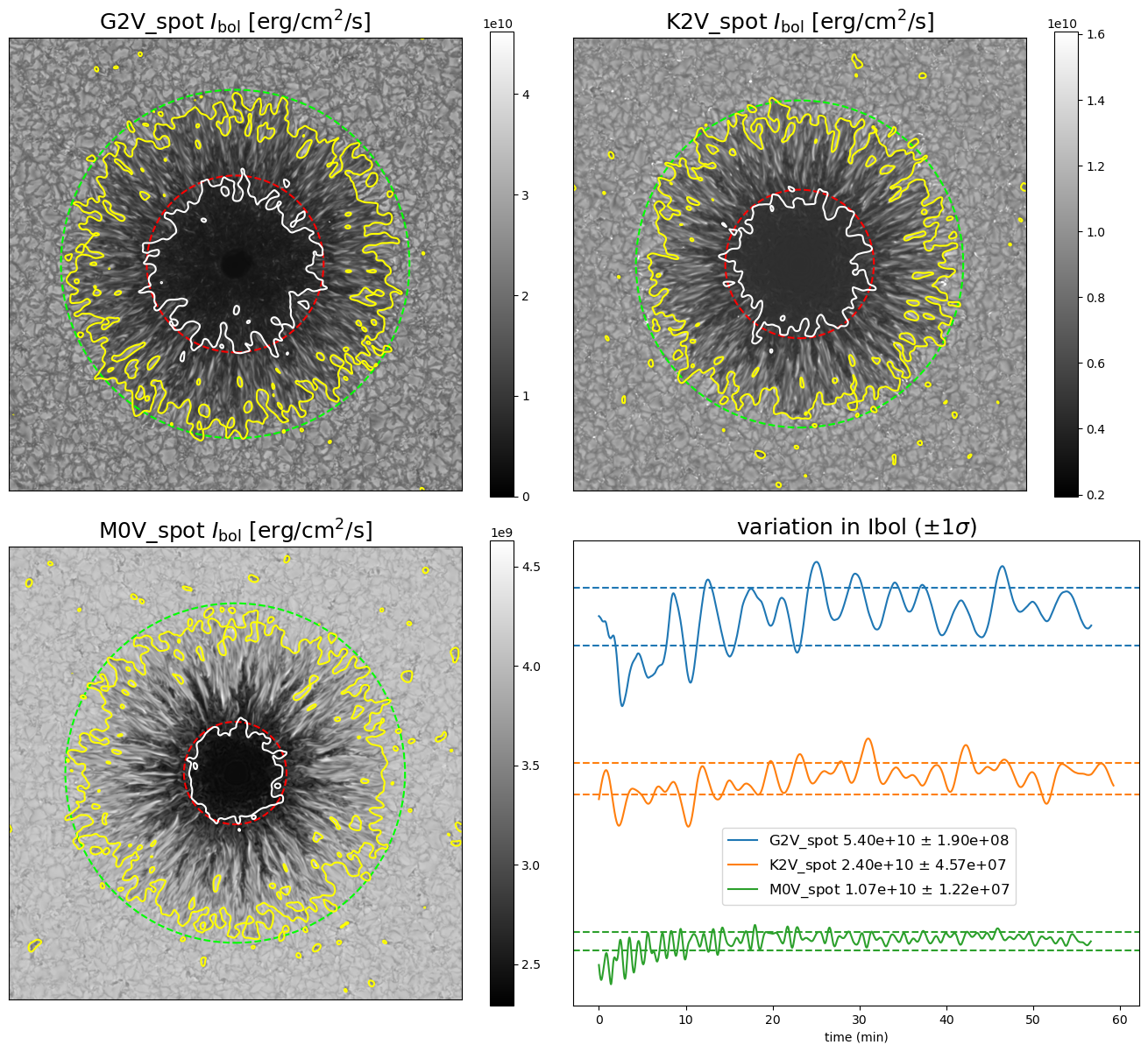}
    \caption{Snapshots of bolometric intensity $I_{\rm bol}$ for G2V (\textit{top left}), K2V (\textit{top right}) and M0V (\textit{bottom left}) spot. The bottom right plot shows the variation in bolometric intensity on an arbitrary axis, with $\pm 1\sigma$ standard deviation horizontal dashed lines. The mean and standard deviation values are mentioned in the figure label. See the corresponding text in Appendix \ref{app:intensity} for details regarding the contours.}
    \label{fig:snap}
    \end{figure}

     The effective temperatures mentioned in Table \ref{tab:setup} are calculated as $T_{\rm eff}=(\pi I_{\rm bol}/\sigma)^{1/4}$. $T_{\rm qs}$ is the quiet star effective temperature and is calculated from mean $I_{\rm bol}$ from a $0.2L_x \times 0.2L_y$ square at the top left corner. $T_{\rm u}$ and $T_{\rm p}$ are the umbra and penumbra effective temperatures calculated from intensity contours, as shown in Figure \ref{fig:snap}. The cut-off values for the umbral and penumbral contours were determined from the intensity histograms and visual inspection such that the visible penumbral filaments are well-covered in the defined penumbral region. The cut-off values finally used are
     \begin{itemize}
         \item G2V: $I_{\rm u}/I_{\rm qs} = 0.35$, $I_{\rm p}/I_{\rm qs} = 0.8$
         \item K2V: $I_{\rm u}/I_{\rm qs} = 0.5$, $I_{\rm p}/I_{\rm qs} = 0.9$
         \item M0V: $I_{\rm u}/I_{\rm qs} = 0.65$, $I_{\rm p}/I_{\rm qs} = 0.965$
     \end{itemize}
    The dashed circles represent the average umbra (red) and spot (green) radius, respectively, as calculated from the area coverage of these contours. This calculation was performed for every snapshot, and then averaged over time to get the radius values used in all the analyses. Lastly, all temperatures, as well as azimuthal averages in Figure \ref{fig:surface} are averaged over an hour of simulation time with a time step of $\sim 6$ minutes.
\end{appendix}

\end{nolinenumbers}
\end{document}